\documentclass{pnastwo}

\usepackage{amssymb,amsfonts,amsmath}
\usepackage{graphics}
\usepackage{graphicx}
\usepackage{setspace}
\usepackage{subfigure}

\contributor{Submitted to Proceedings
of the National Academy of Sciences of the United States of America}
\url{}
\copyrightyear{}
\issuedate{}
\volume{}
\issuenumber{}

\begin{document}



\title{Phase Separation in Hydrogen-Helium Mixtures at Mbar Pressures}




\author{
Miguel A. Morales\affil{1}{Physics Department, University of Illinois at Urbana-Champaign},
Eric Schwegler\affil{2}{Lawrence Livermore National Laboratory},
D. M. Ceperley\affil{1}{Physics Department, University of Illinois at Urbana-Champaign}\affil{3}{National Center of
Supercomputing Applications}\affil{4}{Institute of Condensed Matter Theory, University of Illinois at Urbana-Champaign},
Carlo Pierleoni\affil{5}{CNISM and Physics Department, University of L'Aquila, Via Vetoio, L'Aquila (Italy)}\affil{4}{Institute of Condensed Matter Theory, University of Illinois at Urbana-Champaign},
Sebastien Hamel\affil{2}{Lawrence Livermore National Laboratory}, and
Kyle Caspersen\affil{2}{Lawrence Livermore National Laboratory}
}

\maketitle

\begin{article}

\begin{abstract}
The properties of hydrogen-helium mixtures at Mbar pressures
and intermediate temperatures (4000 to 10000 K) are calculated
with first-principles molecular dynamics simulations. We
determine the equation of state as a function of density,
temperature, and composition and, using thermodynamic
integration, we estimate the Gibbs free energy of mixing,
thereby determining the temperature, at a given pressure, when
helium becomes insoluble in dense metallic hydrogen. These
results are directly relevant to models of the interior
structure and evolution of Jovian planets. We find that the
temperatures for the demixing of helium and hydrogen are
sufficiently high to cross the planetary adiabat of Saturn at
pressures around 5 Mbar; helium is partially miscible
throughout a significant portion of the interior of Saturn, and
to a lesser extent in Jupiter.
\end{abstract}






\dropcap{T}he two lightest elements, hydrogen and helium, are
fascinating to physicists. Ubiquitous in the universe, their
abundance ratio provide stringent checks on cosmological nucleosynthesis
theories and the global distribution of hydrogen in the observable universe provides clues to the origin
and large scale structures of galaxies. They are the essential
elements of stars and giant planets. Yet, in spite of the seeming
simplicity of their electronic structure,
there are many unanswered questions about their fundamental
properties, especially at high pressures. One such question is
under what conditions are these elements miscible. The answer will have a
crucial impact on our understanding of the evolution and the
structure of the giant planets in our solar system and beyond.

Jupiter and Saturn, the simplest among the Jovian planets, are generally believed to have been formed
approximately at the same time as the sun, although certain
direct observations (such as Saturn's excess luminosity) appear to contradict this planetary
formation theory. In addition to being mostly made
of hydrogen and helium, a characteristic of Jovian planets is
that they radiate more energy than they take in from the sun.
Various models of their evolution and structure
have been developed
\cite{Fortney03,Fortney04,Guillot05,Hubbard02} to describe a
relation between the age, volume, and mass of the planet and
its luminosity. The current luminosity of Jupiter is well
described with an evolution model for a convective
homogeneous planet radiating energy left over from its
formation 4.55 billion years ago. But a similar model seriously
underestimates the current luminosity of
Saturn \cite{Hubbard99}. Hence, either Saturn formed much later
than Jupiter, or there is an additional energy source playing a
more important role in Saturn than in Jupiter. In addition, the
atmospheric abundance of helium in both Jupiter and Saturn
appears to be lower than the accepted proto-solar values, more
so in Saturn than in Jupiter \cite{Fortney03}.

Salpeter and Stevenson
\cite{Salpeter73,Stevenson75,Stevenson77_1,Stevenson77_2}
proposed that helium condensation could be responsible for both
the excess luminosity in Saturn and the helium depletion in the
atmosphere of both Jovian planets. Suppose there is a region in
the planet's interior where helium is insoluble; helium
droplets will form and the denser helium will act as a source
of energy, both through the release of latent heat, and by
descending deeper into the center of the planet. Because Jupiter and Saturn
have different total masses, the thermodynamic conditions in
the planetary interiors could be such that this condensation
process is more prevalent in Saturn than in Jupiter.

\begin{figure} 
\includegraphics[scale=0.65]{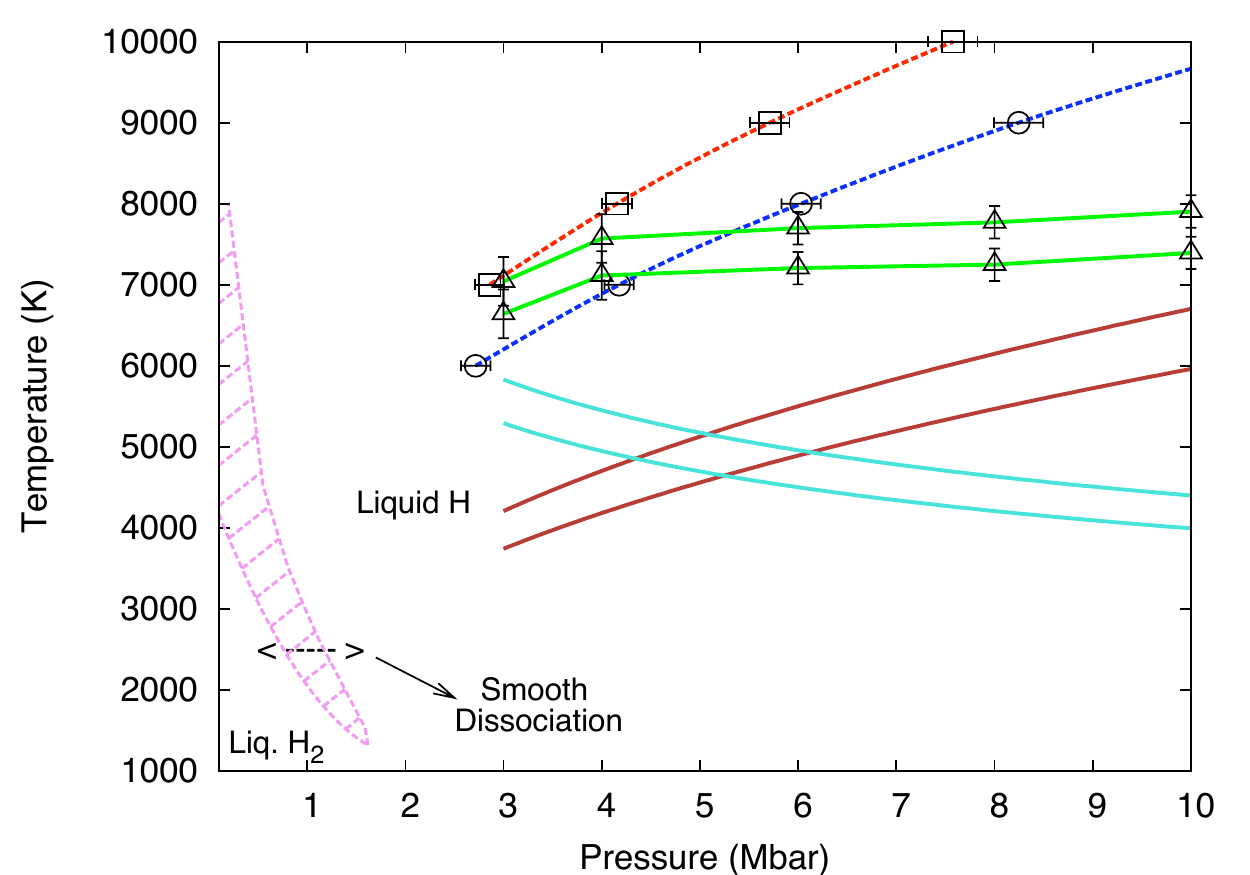}
\caption{\label{full_phase_diagram} Schematic phase diagram of
hydrogen-helium mixtures at two compositions in the relevant
range. Immiscibility lines (solid lines); this work: triangles
(green - guide to the eye), Pfaffenzeller et al.\cite{Pfaff95}
(brown), Hubbard-DeWitt \cite{DeWitt85} (turquoise), the latter
two as parameterized in \cite{Fortney03}. In all three cases,
the upper line corresponds to $x_{He}=0.0847$ and the lower one
to $x_{He}=0.0623$.  Isentropes (dashed lines); Jupiter
assuming a composition $x_{He}=0.07$: squares (red - guide to
the eye), Saturn assuming a composition $x_{He}=0.0667$:
circles (blue - guide to the eye). Also shown in the lower left
corner is the molecular H$_2$ dissociation region of the
mixture (dashed violet) from \cite{Vorberger07} and
\cite{Caspersen08}. }
\end{figure}

Previous attempts to calculate the immiscible temperature, as a
function of pressure and helium concentration, by Stevenson
\cite{Stevenson75}, Straus \cite{Straus77}, Hubbard, {\em et al.}~\cite{DeWitt85}
Klepeis, {\em et al.}~\cite{Klepeis91} and Pfaffenzeller, {\em
et al.}~\cite{Pfaff95} led to inconsistent conclusions as to the importance
of phase separation in the interiors of Saturn and Jupiter.
The original theories of Stevenson, Hubbard and DeWitt were based
on the assumption that the mixture consisted of fully
pressure-ionized hydrogen and helium. For the temperatures and
pressures found in Saturn and Jupiter, this assumption is now
known to be inaccurate, especially for helium
\cite{Stixrude08,Stevenson08}. Klepeis, {\em et al.}~\cite{Klepeis91} and
Pfaffenzeller, {\em et al.}~\cite{Pfaff95} developed mixture
models based on density functional theory (DFT). This opens up
the possibility of providing an accurate description of
electron-ion interactions without assumptions on the extent of
ionization. Klepeis {\em et al.} calculated the enthalpy of
mixing at zero temperature from the analysis of crystal
structures with different concentrations of helium. Using those
enthalpies and the assumption of ideal mixing for the entropy,
they obtained a demixing temperature of 15000 K for
$x_{He}=0.07$, which suggests that there should be a major
phase separation in both Jupiter and Saturn.
However, this work neglected both the relaxation of the ionic
crystal after the introduction of helium, and disorder characteristic of a fluid. Using
first-principles molecular dynamics (FPMD) simulations with the Car-Parrinello technique,
Pfaffenzeller {\em et al.} \cite{Pfaff95} developed a model including
a realistic fluid structure. They performed molecular dynamics (MD) simulations of
fluid pure hydrogen and estimated the free energies of a
mixture by a re-weighting technique. They found a negligible
temperature effect on the mixing free energy up to temperatures
of 3000 K and therefore disregarded thermal effects in
enthalpies of mixing and used the ideal mixing for the entropy.
They obtained immiscibility temperatures too low to allow for
differentiation in either Jupiter or Saturn.


In the present work, the temperature, pressure and composition
dependence of the enthalpy in hydrogen-helium mixtures is
computed with FPMD simulations (see Methods) based on DFT.
DFT has become the method of choice in theoretical studies at high pressures,
producing sufficiently accurate results for hydrogen and helium \cite{Bonev04,Pierleoni08}.
We neglect the zero point energy of the ions, which has been shown to be small
and will have a negligible effect on the immiscibility temperature within our precision \cite{Pierleoni04}.
Using thermodynamic integration, we estimate the
Helmholtz free energy of the mixture and determine the demixing
temperature as a function of pressure and composition, thereby avoiding many of the
previous assumptions and providing the most accurate prediction
of the hydrogen-helium immiscibility, to date. Figure \ref{full_phase_diagram}
summarizes the main findings of this work. The isentropes for
Jupiter and Saturn determined from our DFT-based equation of
state (EOS) are shown along with the temperature of demixing.
Overall, we find that the demixing temperature is high enough
to support the scenario where helium is partially miscible over
a significant fraction of the interior of the Jovian planets,
with the corresponding region in Saturn being larger than in
Jupiter.

\section{Results and Discussion}

We calculated the EOS of the hydrogen-helium system as a
function of composition in the temperature range 4000 to 10000 K
and in the density range 0.3 to 2.7 g/cm$^3$, by a series of
FPMD simulations in the NVT ensemble. We studied 12 different
compositions to obtain an accurate interpolation of
the energy and pressure. Using the EOS we calculated free
energies by integrating along isotherms and isochores. We used
the following multistep process to estimate the Gibbs free
energy:
\begin{enumerate}
\item We computed the free energy of an
effective model at the ``reference point''
($T_{ref} = 10,000K$, $r_{s}$ = 1.25) \footnote{ $r_{s}$ is the
Wigner-Seitz radius and defines the electronic density.}  for all 12
compositions (see Methods section). At this point the structure of the liquid mixture can be reasonably well
reproduced by simple pair potentials between classical point
particles.
\item Using Coupling Constant integration (CCI) (see Methods section), we computed the
free energy difference at the reference point between the DFT
model and the effective model.
\item Integrating the EOS along constant temperature and constant volume paths, we obtained the
free energy difference between the reference point and any other thermodynamic point
in the range investigated.
\item Finally, inverting the pressure-volume relations we  obtained the Gibbs free energy of
mixing as a function of pressure, temperature and composition.
\end{enumerate}

\begin{figure} 
\includegraphics[scale=0.65]{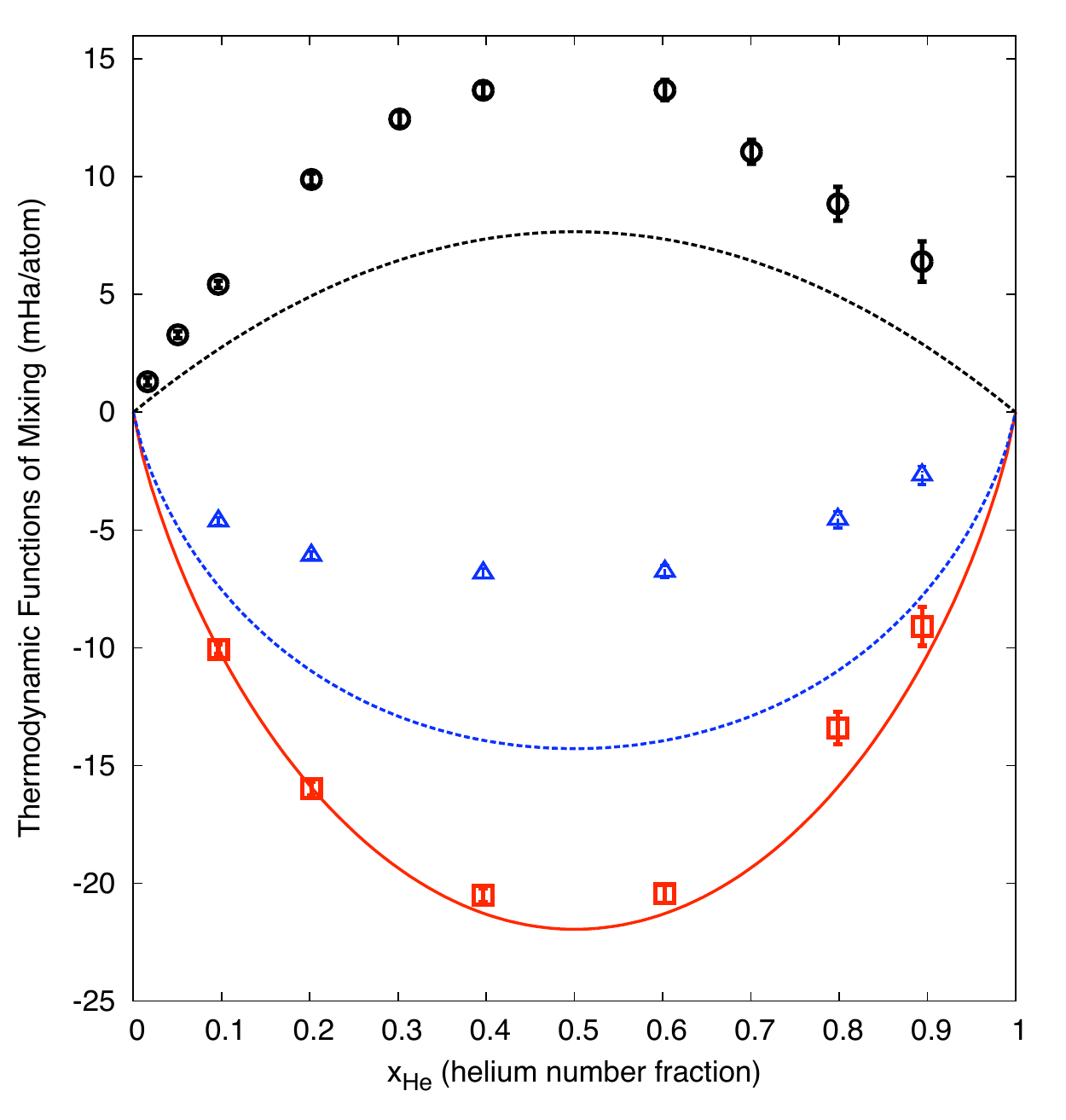}
\caption{\label{refF} Internal energy ($\circ$ black), entropy contribution to the free energy ($\square$ red)
and Helmholtz free energy ($\triangle$ blue) of mixing
as a function of the helium number fraction, at the
reference point. The red solid line represents the ideal entropy of mixing
and the black and blue dashed lines are results from Pfaffenzeller et al.\cite{Pfaff95}. }
\end{figure}

Figure \ref{refF} shows the Helmholtz free energy of mixing
as a function of helium number fraction ($x_{He}$) at the
reference point as well as its energetic and entropic
contributions. The figure clearly shows, at least at the reference point,
that the ideal mixing assumption to describe the entropy
of mixing is very accurate for $x_{He} \leq 0.2$ but becomes
less accurate for larger helium concentrations.
The reason for this behavior is that the local environment of a proton
in the low hydrogen concentration region is very different from
the one it experiences in the metallic state of the pure system (see the discussion
of pair correlations below). On the other hand, the inert character of helium makes it insensitive
to change in the local environment in the low helium concentration region.
In figure \ref{refF} we compare our results to the prediction of
Pfaffenzeller et al. \cite{Pfaff95} who neglected thermal
effects in the internal energy of mixing and used the ideal
mixing law. The neglect of thermal effects in
the internal energy results in a too large and negative mixing
free energy. While it is true that the thermal effects are
probably negligible at 3000 K, at this temperature the
system is strongly immiscible so that the re-weighting
procedure used in \cite{Pfaff95} to estimate those effects
is likely to be inaccurate.

\begin{figure}
\includegraphics[scale=0.6]{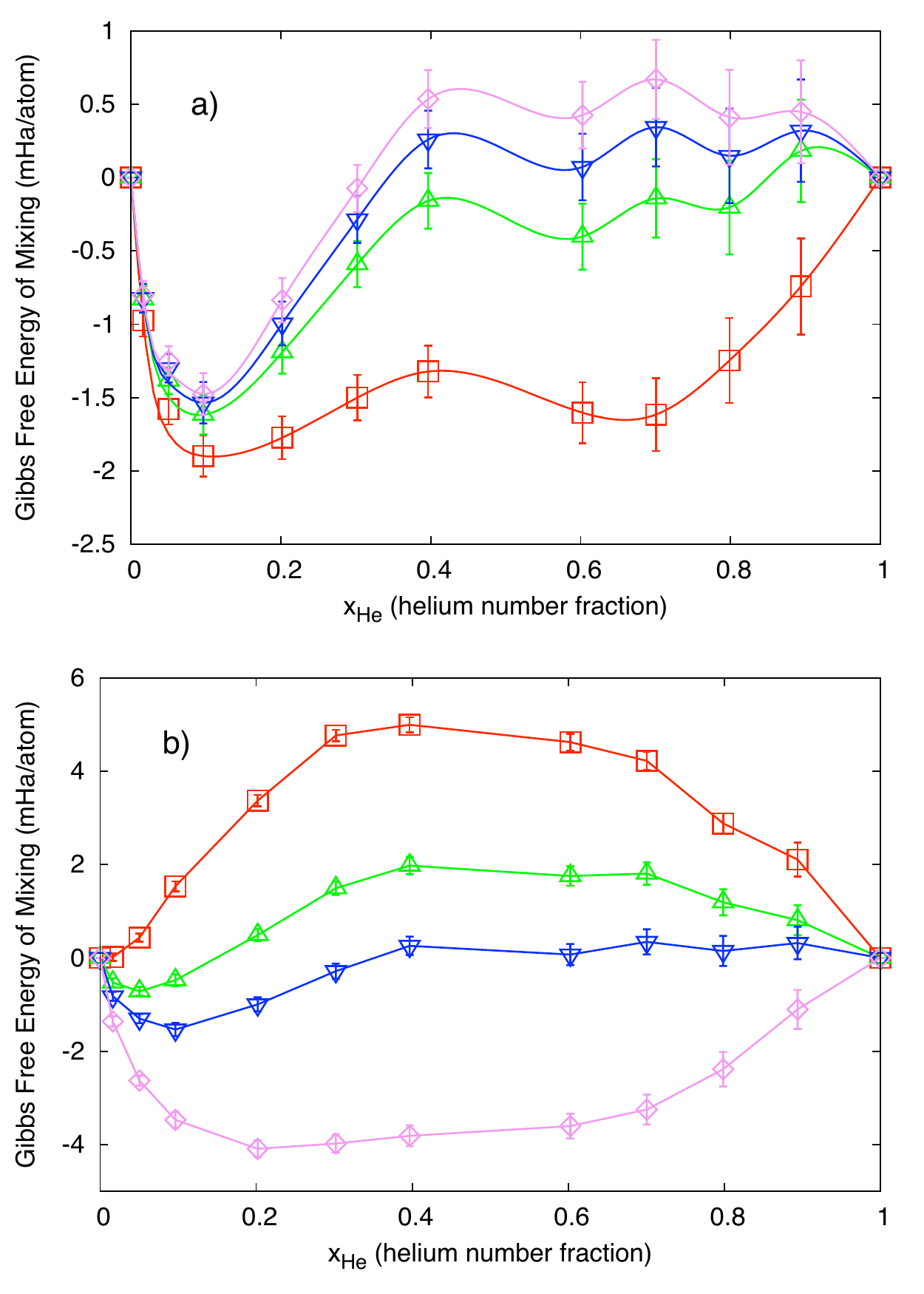}
\caption{\label{gm_T} The Gibbs free energy of mixing as a
function of composition. Panel a), Data at 8000K for several pressures: 4 Mbar ($\square$ - red), 8
Mbar ($\triangle$ - green), 10 Mbar ($\triangledown$ - blue) and 12 Mbar($\diamond$ - violet). Panel b), Data at P= 10 Mbar for several temperatures: 5000K ($\square$ -red), 7000K ($\triangle$ - green), 8000K ( $\triangledown$ - blue) and 10000K ($\diamond$ - violet). Results of calculations are shown as symbols and error bars; solid lines are guides to the eye only. }
\end{figure}

In figure \ref{gm_T} we
present the calculated Gibbs free energy of mixing as a
function of composition; in panel a) several pressures are
shown at a temperature of 8000 K, while in panel b) several
temperatures are shown at a pressure of 10 Mbar. Note that at
8000 K, pressure has a small effect on the mixing free energy
for low helium concentrations. In particular, a minimum in the
free energy located at $x_{He}\sim 0.1$ is observed for all
pressures investigated; this implies a stable mixture at this
concentration.  On the other hand, pressure has a strong effect
at higher helium concentrations where, at a temperature of 8000
K, an increase from 4 to 10 Mbar eliminates the second minimum in the
free energy curve.\footnote{We have concentrated the majority of our simulation
efforts on the small $x_{He}$ part of the phase diagram more
relevant to planetary models. Quantitative prediction of
miscibility at large $x_{He}$ is more difficult because of the
smaller mixing free energy involved and will require additional
investigations.} The common tangent construction is used to estimate the demixing
temperatures. For points where no minima at high helium
concentration is evident, we have assumed complete
immiscibility. From the free energy plots of figure \ref{gm_T}
it is clear that this assumption will have
a negligible effect on the location of the minimum at low
helium concentration. As shown in panel b) of figure \ref{gm_T}, temperature has a
strong effect on the mixing free energy, and hence, on
immiscibility. An increase in temperature from 7000K to 9000K (not shown in the figure) is enough to change the concentration of helium at the saturation point from $5\%$ to
$15\%$.

\begin{figure}
\includegraphics[scale=0.62]{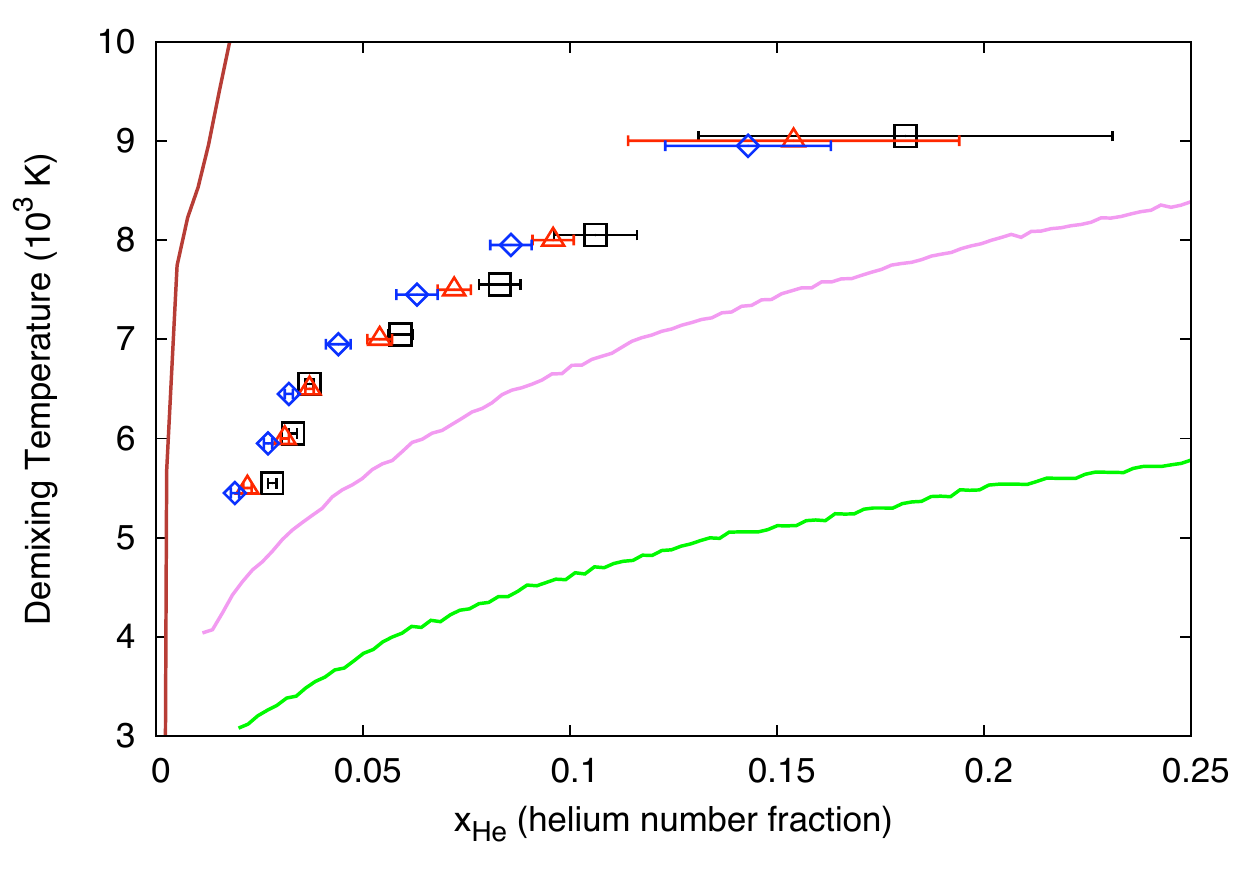}
\caption{\label{critT} Demixing transition temperatures as a
function of helium number fraction, for several pressures:  4
Mbar ($\square$, black), 8 Mbar ($\triangle$, red), 12
Mbar ($\diamond$, blue). Continuous lines are the results of
Ref.~\cite{Klepeis91} for 10.5 Mbar (brown) and
Ref.~\cite{Pfaff95} for 8 Mbar (green) and 10 Mbar (violet). }
\end{figure}

Figure \ref{critT} shows the demixing temperature versus
composition for pressures ranging from 4 to 12 Mbar. Also
shown are the results from the previous DFT-based calculations
\cite{Klepeis91,Pfaff95}. As suggested by the free energy
curve in figure \ref{gm_T}, pressure has only a moderate effect
on the immiscibility process. For a fixed helium fraction, the
demixing temperature changes by approximately 500 K in a
pressure range of 8 Mbar for the relevant concentrations
($5\%$ to $10\%$).

Recent first-principles studies of pure helium have examined
the effect of temperature on band gap closure, suggesting that
metallization in helium can occur at much smaller pressures
than previously expected \cite{Stixrude08}. To examine the
nature of helium in the mixtures, we calculated the electronic
conductivity of pure helium using the Kubo-Greenwood approach
within DFT\footnote{These calculations were performed on 15
de-correlated configurations from a FPMD trajectory, where we
employed a 6$\times$6$\times$6 Monkhorst-Pack k-point grid.}
and obtained values well below 100 $(\Omega cm)^{-1}$, even at
the highest temperature and density reported here. Furthermore,
in the recent work by Stixrude \emph{et al.} \cite{Stixrude08},
for $\rho \approx$ 5.4 $g/cm^3$ the band gap is found to close
at temperatures beyond 20000 K, well above our estimated
demixing temperature. Metallization should enhance helium
solubility, but as clearly shown here, for the pressures
relevant to the modeling of Jovian planets, immiscibility
occurs at temperatures well below those required to produce
ionization in helium \cite{Stevenson08}; fully ionized models are not appropriate
for describing the pressure dependence of the demixing
temperature. At pressures much higher than those examined here,
metallization of helium will play an important role and should
produce significant changes to the pressure dependence of the
immiscibility temperature.

\begin{figure} 
\includegraphics[scale=0.68]{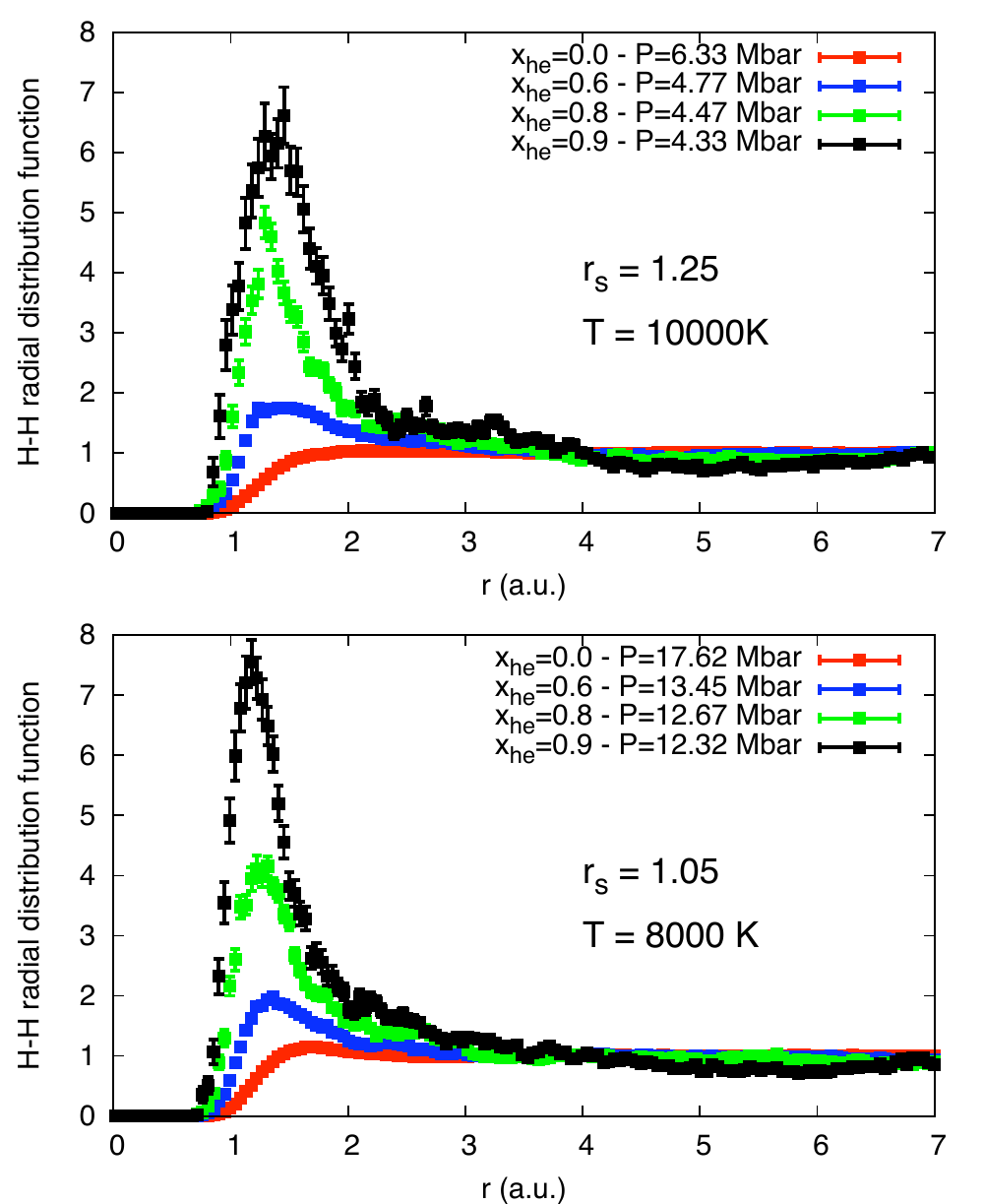}
\caption{\label{grs} The effect of increasing concentrations of
helium on the proton-proton radial distribution function at $r_s =
1.25$ (top panel) and $r_s = 1.05$ (bottom panel). }
\end{figure}

The structure of hydrogen is strongly influenced by the helium concentration.
While at low $x_{He}$ hydrogen is in the mono-atomic fully ionized state,
an effective proton-proton attraction reminiscent of the molecular bonding
develops upon increasing $x_{He}$, even at very
high pressures and temperatures. Figure \ref{grs} shows several hydrogen-hydrogen
radial distribution functions for mixtures with various helium
concentrations, for temperatures of 8000 K and 10000 K and
electronic densities given by $r_s = 1.05$ and $r_s = 1.25$
respectively. A molecular-like peak builds up smoothly as
$x_{He} \rightarrow 1 $.  Under these conditions, helium is not
ionized; this inhibits the delocalization of the hydrogenic
electrons, enhancing the formation of weak molecular bonds.
Because of the very low proton concentration, the observed
proton-proton correlation can be interpreted as resulting from
an effective Morse potential.  Fitting $-\log(g_{pp}(r))/T$ to
this analytic form yields well depth parameters $\sim$300 times
smaller than in an isolated hydrogen molecule. Such weak
attraction gives proton pairs with short lifetimes, as also
inferred from direct inspection of the MD trajectories.
A similar stabilization of molecular
hydrogen by helium, but at much lower temperature and density, has been previously reported
close to the dissociation regime in pure hydrogen
\cite{Vorberger07}.

The computed demixing temperatures found here have important
implications for the study of the interior structure of
hydrogen rich planets, especially Saturn. Our results support
the scenario where helium becomes partially miscible in the
intermediate layers of the planet, with the excess helium
falling towards the core through  gravitational
differentiation. This mechanism has been proposed to explain
the high surface temperatures observed in Saturn
\cite{Fortney04} and the depletion of helium in the atmosphere.
Whether the immiscible region will be large enough to explain
all the observed properties of the planet remains an open
question, but the current work represents a clear indication
that a correct description should include phase separation. In
general, DFT represents a large modification to the existing
interior models, based mostly on the Saumon, Chabrier, Van Horn
EOS \cite{Saumon95} (SCVH). As recently shown by Militzer \emph{et al.} \cite{Militzer08},
a combination of the DFT EOS with the
inclusion of non-ideal effects in the mixing leads to
isentropic Jupiter models that are cooler than the
corresponding SCVH ones by about 1000-2000 K at high pressures.
Even though the results reported here suggest that the interior
model can not be assumed to be isentropic, non-adiabatic models
based on DFT including non-ideal effects should produce
qualitatively similar results to those found by Militzer. In
order to give an indication of the general changes expected
with the proper inclusion of these effects, we calculated the
SCVH isentropes (corresponding to the interpolated EOS) of
Jupiter and Saturn using our DFT-based EOS. The results are
shown in figure~\ref{full_phase_diagram}, where we have assumed
fixed helium concentrations of $x_{He}=0.07$ for Jupiter and
$x_{He}=0.0667$ for Saturn.

At low density ($\rho \leq 0.3$ $g/cm^3$), before the
dissociation of  hydrogen plays a significant role, the
pressures produced by the SCVH model  are in good agreement
with DFT. In order to compare the entropy of both models at low
density, and assess the agreement of the isentropes there, we
performed simulations with $x_{He}=0.0667$ following a path
through the region of miscibility starting from the reference
point and ending at a density of $\rho = 0.3$ $g/cm^3$, where
we simulated several temperatures. Then using thermodynamic
integration, we calculated the free energy and the entropy at
the point where the SCVH planetary isentrope crosses this
density. We obtain entropies that agree with the SCVH model to
within 1.5$\%$. With the onset of dissociation for higher densities,
the isentrope from the SCVH model deviates significantly from DFT
results. Although the possible immiscibility puts in question the use
of an isentropic model for the interior of Saturn, our analysis provides
 an estimate of the magnitude of changes to the SCVH based models.
 The present results are in good agreement with those reported by Militzer \cite{Militzer08}
for the Jupiter isentrope. As can be seen, the portion of the
interior of Saturn corresponding to pressures between 1.5 to 5.5
Mbar should be inside the immiscible region, with the stable
concentration of helium depending on pressure.

\section{Conclusion}

In summary, we have carried out an extensive investigation of
the properties of hydrogen-helium mixtures at pressures and
temperatures that are relevant to the interior of Jupiter and
Saturn using state-of-the-art {\it{ab initio}} simulation methods.
By using a combination of first-principles molecular
dynamics simulations within DFT and thermodynamic integration
techniques, we have accurately determined the Gibbs free energy
of mixing over a wide range of density, temperature and
composition. Our work differs from previous investigations in that it does not rely on
any assumptions about mixing functions. Our simulation results are consistent with the
idea that a large portion of the interior of Saturn has
conditions such that hydrogen and helium phase separate; this
can account for the apparent discrepancy between the current
evolution models for Saturn and observational data.

The accuracy of the present results is primarily limited
by the approximate density functional used. Quantum Monte Carlo calculations
for helium and hydrogen do not require assumptions about the electron correlation,
pseudopotentials and zero-point energy of the nuclei. A systematic investigation of the
correction to DFT using the Quantum Monte Carlo method described in \cite{Pierleoni08} is in progress.


\section{Methods}

\subsection{First Principles Calculations}
The FPMD simulations performed in this work were based on
Kohn-Sham density functional theory using the
Perdew-Burke-Ernzerhof (PBE) exchange-correlation functional.
We employed Born-Oppenheimer  MD (BOMD) within the NVT-ensemble
(with a weakly coupled Berendsen thermostat), as implemented in
the Qbox code (http://eslab.ucdavis.edu/software/qbox). We used
a Hamman type \cite{Hamman89} local pseudopotential with a core
radius of $r_c=0.3$ au to represent hydrogen and a
Troullier-Martins type \cite{Troullier91} nonlocal
pseudopotential with $s$ and $p$ channels and $r_c=1.091$ au to
represent helium. Tests have established the accuracy of the
hydrogen and helium pseudopotentials over the relevant pressure
range, up to 13 Mbars. \footnote{The sensitivity of the free
energy to the choice of the helium pseudopotential was
estimated by comparing the enthalpy of 20 ionic configurations
with a Hamman type local pseudopotential with $r_c=0.218$ au
for helium and a plane wave cutoff of 450 Ryd. The difference
in the free energy of mixing is estimated to be 0.1 mH/atom,
this is smaller than the error bars reported in
Fig.~\ref{gm_T}.} A plane wave energy cutoff of 90 Ry was used
for $r_s \geq 1.10$  and of 115 Ry for $r_s < 1.10$. Empty
states were included with an electronic temperature set to the
ionic temperature. To integrate the equation of motion during
the dynamics we used a time step of 8 a.u.

We used 250 electrons for $r_s < 1.65 $  and 128 electrons
otherwise. The Brillouin zone was sampled at the
$\Gamma$-point. In order to reduce systematic effects and to
get accurate pressures, we added a correction to the EOS
designed to correct for the plane wave cutoff and the sampling
of the Brillouin zone. To compute this, we studied 15-20
configurations at each density and composition by using a 4x4x4
grid of k-points with a plane-wave cutoff of at least 300 Ry.
The actual plane-wave cutoff used depended on density and
was chosen to achieve full convergence in the energy and pressure.

For the calculation of the EOS, we studied 4 temperatures, 6
densities and 12 compositions for a total of 288 simulations.
We also performed 15 simulations to extend the calculation of
the free energy to low density. For the integration of the free energy
we studied 8 compositions, each one required 5 additional calculations.
In the case of the BOMD simulations, we first equilibrated the system
using a suitable effective model and subsequently allowed 300-500 timesteps
of equilibration with DFT, averages were accumulated for approximately 2000 time steps.
For the thermodynamic integration, the simulations were first equilibrated with the effective potential
and subsequently allowed to equilibrate for 1000-2000 timesteps
using the mixed DFT-effective potential, averages were
calculated for 12000-18000 timesteps. In both cases, this was sufficient for accurate results.
In total, the simulations reported here used approximately three million CPU hours on a
large Opteron-based Linux cluster.

\subsection{Coupling Constant Integration}
CCI allows us to calculate the difference in free energy
between systems with different interacting potentials. For a
system described by the potentials $V_1$ and $V_2$:
\begin{eqnarray}
\label{pot1}
V(\lambda) & = & \lambda V_1 + (1-\lambda) V_2  \\
\label{pot2}
F_1(T,V,N) - F_2(T,V,N) & = & \int_0^1 \frac{dF(\lambda)}{d\lambda} d\lambda \nonumber \\
& = & \int_0^1 \left < V_1 - V_2 \right >_{T,V,N,\lambda}
d\lambda, \ \ \
\end{eqnarray}
where $ \left < \right >_{T,V,N,\lambda}$ represents a
canonical average with the potential $V(\lambda)$. Any
functional form of the two potentials is formally allowed in
Eq.~\ref{pot1}, but the use of similar potentials makes the
integration of the free energy difference considerably easier
in practice.

\begin{figure}
\includegraphics[scale=0.5]{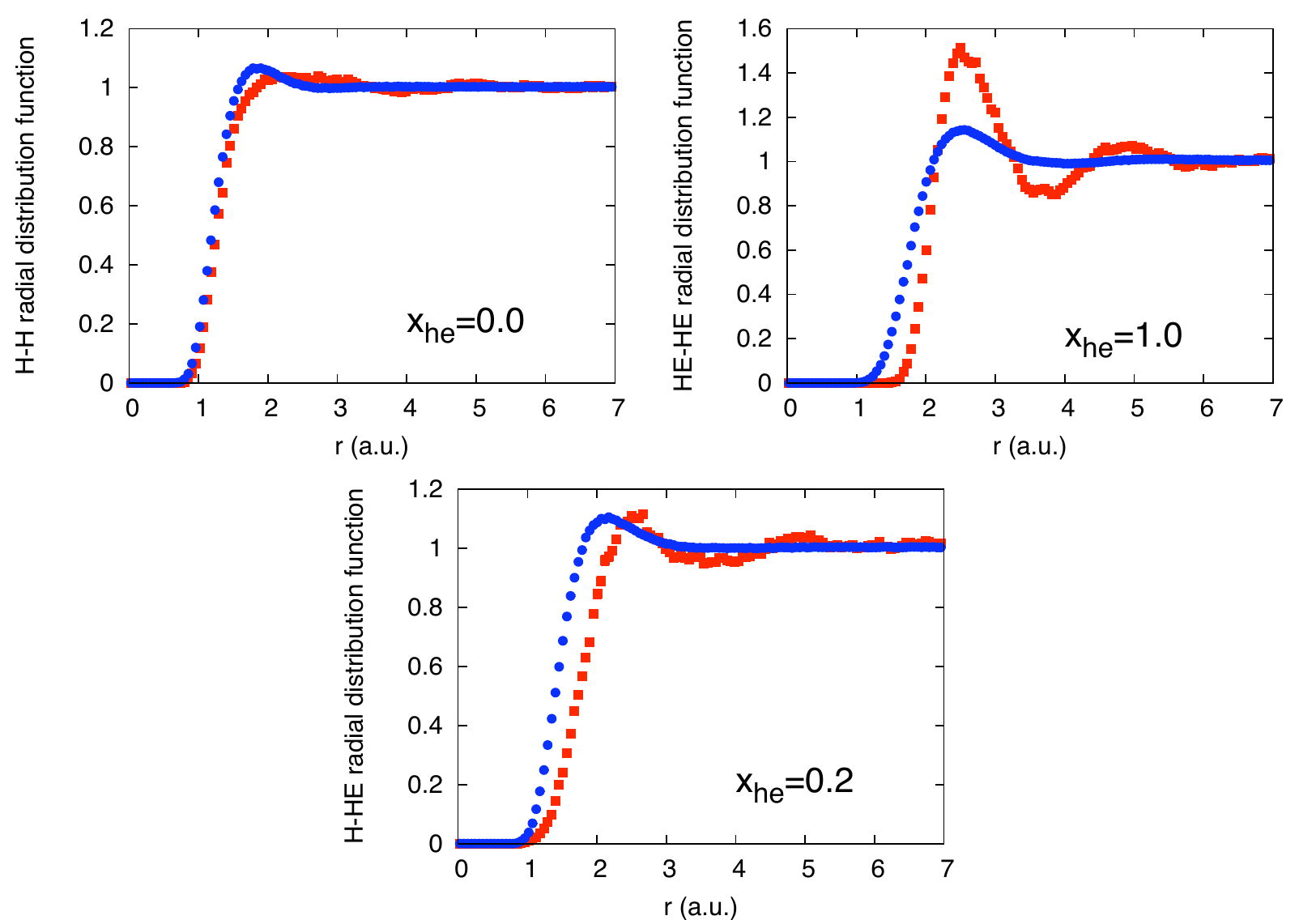}
\caption{\label{ccti}  Comparison of several radial distribution
functions between the effective (blue) and the DFT (red) system
at the reference point. We choose Yukawa pair potentials to
calculate the Helmholtz free energy at the reference point.}
\end{figure}

To represent the interaction between the atoms in the classical
system, we used reflected Yukawa pair potentials:
\begin{equation}
V(r) =
\begin{cases}
a ( \frac{e^{-br}}{r} + \frac{e^{-b(L-r)}}{(L-r)} - 4 \frac{e^{-bL/2}}{L}) & r \le L/2, \\
0 & r > L/2,
\end{cases}
\end{equation}
where a,b and L are free parameters and depend on the identity of
the atoms\footnote{We used: $a_{H}$=$a_{He}$=$a_{H-He}$=1, $L_{H}$=$L_{He}$=$L_{H-He}$=8 a.u., $b_{H}$= 2.5 a.u., $b_{He}$=1.2 a.u. and $b_{H-He}$=1.9 a.u.}. As shown in figure \ref{ccti},
this potential was found to exhibit similar pair correlations as the
DFT model for pure hydrogen, but was not as good for helium.
We choose the pair potentials such that the effective model was
fully miscible at the reference point to avoid crossing a phase
line during the integration. We computed the Helmholtz free
energy of the effective model using CCI and classical MC
simulations, with the second potential set to zero in Eq.
\ref{pot1}.  From this we determined the free energy of the
effective model since the free energy of the non-interacting
model is known. We also calculated the free energy by
integrating the pressure from the reference volume to a volume
large enough that the system is ideal; the pressures were
obtained by a series of classical MC simulations. Both
approaches produced agreement within noise.

\begin{figure}
\includegraphics[scale=0.45]{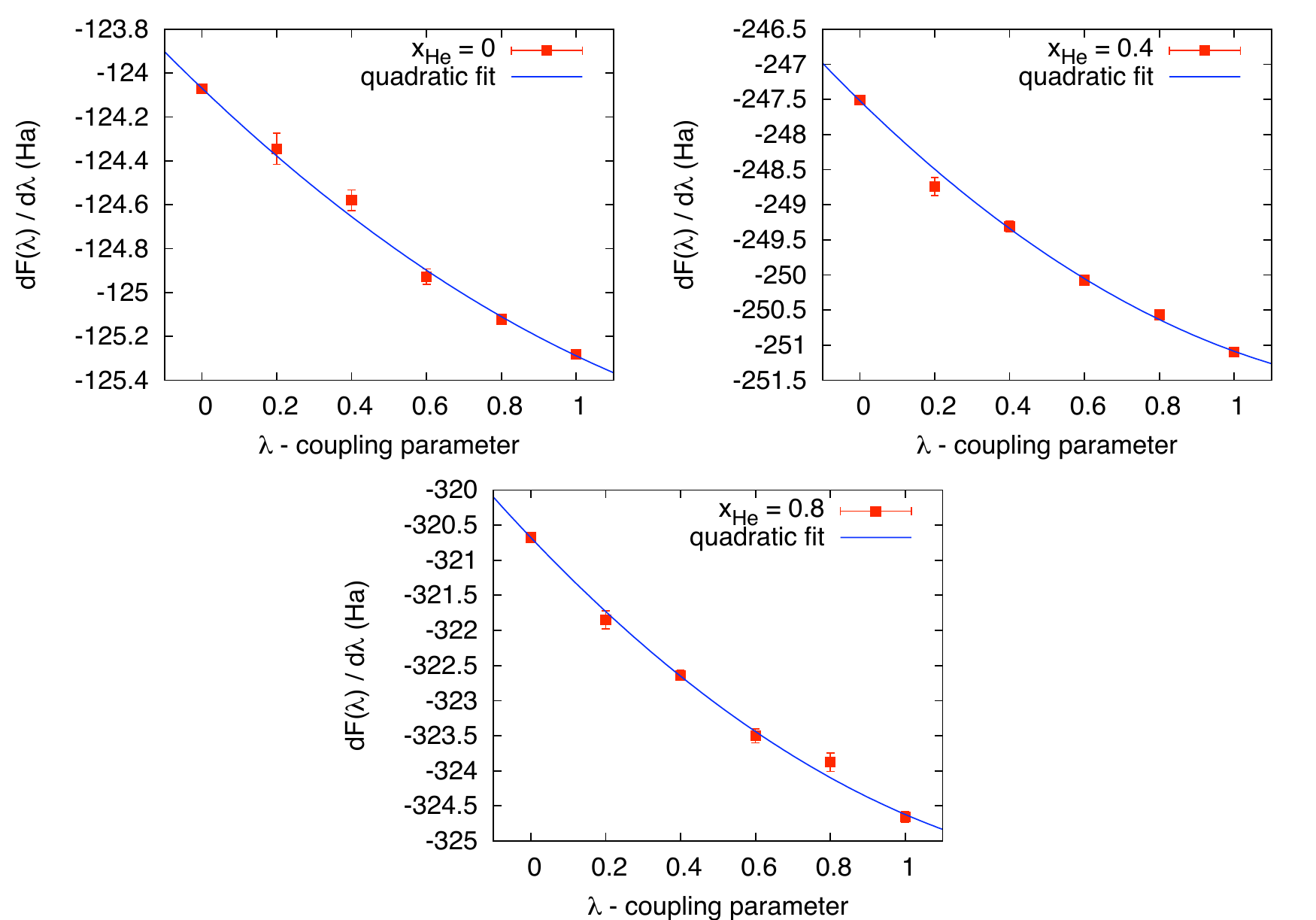}
\caption{\label{ccti_dftcl}  Results of the simulations for the
mixed effective-DFT potential used in the CCI of the free
energy difference for several compositions. As can be seen, the
results depend smoothly on the coupling parameter and no
singular behavior is observed close to the endpoints.}
\end{figure}

The free energy difference between the DFT-based and the
effective models was calculated using the CCI approach. Figure
\ref{ccti} shows a comparison of the radial distribution functions
of the effective and DFT models for selected compositions at
the reference point. To compute the required canonical
averages, we used a Hybrid Monte Carlo (HMC) algorithm
\cite{Mehlig92}, which allows for an efficient sampling of
large systems with many-particle moves. HMC results in exact
sampling of the canonical ensemble without time step errors. In
this work, the HMC approach was found to be as efficient as MD
if the time step is chosen carefully. Figure \ref{ccti_dftcl}
shows the results of the HMC simulations for several
compositions. The curves are smooth. This is the only requirement to justify
the procedure. Figure \ref{refF} summarizes the main results of
the computations at the reference point.

\begin{acknowledgments}
We acknowledge useful comments from D.J. Stevenson and N.W.
Ashcroft. MAM and DMC acknowledge support from DOE-NNSA and
DOE-DE-FG52-06NA26170. CP acknowledges the support from the
Institute of Condensed Matter Theory at the University of
Illinois at Urbana-Champaign for a short term visit and
financial support from the MIUR-PRIN2007. Extensive
computational support was provided by the Livermore Computing
facility. This work was partly performed under the auspices of
the U.S. Department of Energy by Lawrence Livermore National
Laboratory under Contract DE-AC52-07NA27344.
\end{acknowledgments}


\begin{thebibliography} {100}

\bibitem{Fortney03} Fortney JJ, Hubbard WB (2003) Phase separation in giant planets: inhomogeneous evolution of Saturn. Icarus 164:228-243.

\bibitem{Fortney04} Fortney JJ, Hubbard WB (2004) Effects of helium phase separation on the evolution of extrasolar giant planets. Astrophys J 608:1039-1049.

\bibitem{Guillot05} Guillot T (2005) The interiors of giant planets: Models and outstanding questions. Annu Rev Earth Planet Sci 33:493-530.

\bibitem{Hubbard02} Hubbard WB, Burrows A, Lunine JI (2002) Theory of giant planets. Annu Rev Astron Astrophys 40:103-136.

\bibitem{Hubbard99} Hubbard WB, et.al (1999) Comparative evolution of Jupiter and Saturn. Planet. Space Sci. 47:1175-1182.

\bibitem{Salpeter73} Salpeter EE (1973) On convection and gravitational layering in Jupiter and in stars of low mass. Astrophys J 181:L83-L86

\bibitem{Stevenson75} Stevenson DJ (1975) Thermodynamics and phase separation of dense fully ionized hydrogenÐhelium fluid mixtures. Phys Rev B 12:3999-4007.

\bibitem{Stevenson77_1} Stevenson DJ, Salpeter EE (1977) Phase diagram and transport properties for hydrogen-helium fluid planets. Astrophys J Suppl 35:221-237.

\bibitem{Stevenson77_2} Stevenson DJ, Salpeter EE (1977) Dynamics and helium distribution in hydrogen-helium fluid planets. Astrophys J Suppl 35:239-261.

\bibitem{Straus77} Straus DM, Ashcroft NW, Beck H (1977) Phase separation of metallic hydrogen-helium alloys. Phys. Rev. B 15:1914-1928.

\bibitem{DeWitt85} Hubbard WB, DeWitt HE (1985) Statistical mechanics of light elements at high pressure. VII - A perturbative free energy for arbitrary mixtures of H and He. Astrophys. J. 290:388-393.
\bibitem{Klepeis91} Klepeis JE, Schafer KJ, Barbee TW, Ross M (1991) HydrogenÐ helium mixtures at megabar pressures - Implications for Jupiter and Saturn. Science 254:986-989.

\bibitem{Pfaff95} Pfaffenzeller O, Hohl D, Ballone P (1995) Miscibility of hydrogen and helium under astrophysical conditions. Phys Rev Lett 74:2599-2602.

\bibitem{Stixrude08} Stixrude L, Jeanloz R (2008) Fluid helium at conditions of giant planetary interiors, Proc Natl Acad Sci USA 105:11071-11075.

\bibitem{Stevenson08} Stevenson DJ (2008) Metallic helium in massive planets, Proc Natl Acad Sci USA 105:11035-11036.


\bibitem{Bonev04} Bonev SA, Schwegler E, Ogitsu, Galli G (2004) A quantum fluid of metallic hydrogen suggested by first-principles calculations. Nature 431:669-672.

\bibitem{Pierleoni08} Pierleoni C, Delaney KT, Morales MA, Ceperley DM, Holzmann M (2008) Trial wave functions for high pressure metallic hydrogen. Comput Phys Commun 179:89-97.

\bibitem{Pierleoni04} Pierleoni C, Ceperley DM, Holzmann M (2004) Coupled Electron-Ion Monte Carlo calculations of dense metallic hydrogen. Phys Rev Lett 93:146402:1-4.
\bibitem{Vorberger07} Vorberger J, Tamblyn I, Militzer B, Bonev SA (2007) Hydrogen-helium mixtures in the interiors of giant planets. Phys Rev B 75:024206.

\bibitem{Caspersen08} K. Caspersen; S. Hamel, T. Ogitsu, F, Gygi and E.R Schwegler, to be published.

\bibitem{Saumon95} Saumon D, Chabrier G, Van Horn HM (1995) An equation of state for low-mass stars and giant planets. Astrophys J Suppl S 99:713-741.



\bibitem{Militzer08} Militzer B, Hubbard WB, Vorberger J, Tamblyn I, Bonev SA (2008) A massive core in jupiter predicted from first-principles simulations. Astrophys J Lett., in press

\bibitem{Hamman89} Hamman, D.R. (1989) Generalized norm-conserving pseudopotentials. Phys. Rev. B 40: 2980-2987.

\bibitem{Troullier91} Troullier, N. and Martins, J.L. (1991) Efficient pseudopotentials for plane-wave calculations. Phys. Rev. B 43:1993-2006.


\bibitem{Mehlig92} Mehlig B, Heermann DW, Forrest BM (1992) Hybrid Monte Carlo method for condensed-matter systems. Phys Rev B 45:679-685.



\end{thebibliography}





\end{article}








\end{document}